\documentclass[orivec]{llncs}

\usepackage[style=alphabetic,backend=bibtex,isbn=false]{biblatex}
\usepackage{./lib/kbibs/bibtweaks}
\usepackage{graphicx,url,wrapfig,amsmath,amssymb}

\usepackage{subcaption}
\usepackage[show]{ed} 
\usepackage{xcolor}
\usepackage{tabularx}
\usepackage{macros/basics}
\usepackage{comment}
\def\myparagraph#1{\noindent\textbf{\large #1}}

\begin{document}
\title{TGView3D System Description: 3-Dimensional Visualization of Theory Graphs}
\author{Richard Marcus\inst{1} \and Michael Kohlhase\inst{1} \and Florian Rabe\inst{1,2}}
\institute{FAU Erlangen-Nuremberg \and LRI Paris}
\maketitle

\begin{abstract}
  We describe TGView3D, an interactive 3D graph viewer optimized for exploring theory graphs.
  To exploit the three spatial dimensions, it extends a force-directed layout with a hierarchical component.
  Because of the limitations of regular displays, the system also supports the use of a head-mounted display and utilizes several virtual realty interaction concepts.
\end{abstract}

\myparagraph{Introduction}
Libraries of both informal and formal mathematics have reached enormous sizes: at least half-a-dozen of them exceed $10^5$ statements.
Thus, it is getting more and more difficult to organize this knowledge in a way that humans can understand and therefore access it. 
This can lead to the duplication of formalizations because users are unaware of them or cannot find them.
Often systems employ modularization features to introduce a high-level structure that helps to navigate through the libraries.
Many of these features can be captured in the language of theories and theory morphisms as used in, e.g., the OMDoc/MMT format \cite{Kohlhase:OMDoc1.2,RabKoh:WSMSML13}, which has been used as a uniform representation standard for libraries related to theorem proving, computation, and mathematical documents \cite{KohRab:qrtpflmk15}.

To help users understand this high-level structure better, we have experimented with visualizing theory graphs~\cite{Kohlhase:mkmtobbtg14} in the interactive 2D graph viewer TGView~\cite{RupKohMue:fitgv17}.
But the sheer size of theory graphs --- they can easily contain thousands of nodes and an order of magnitude more edges --- complicates the process of computing structured graph layouts.
We have identified the use of three dimensions as a possible solution as this offers more space to organize the theory graph.
Extending this idea, we have integrated virtual reality devices to allow a wide field of view and intuitive ways of interacting with the 3D world.

\paragraph{Contribution}
Consequently, we have developed a virtual reality graph viewer, TGView3D, which we describe in this paper.
We use the \textbf{Unity} game engine as a basis because it offers integration for popular VR-devices and supports many platforms.
The system is licensed under GPLv3 and is is available at \url{https://github.com/UniFormal/TGView3D}.
A demo video can be found at at \url{https://www.youtube.com/watch?v=Mx7HSWD5dwg}.
A preliminary demonstration was presented at CICM 2018 as a demo-only presentation, i.e., without any accompanying write-up.

\paragraph{Related Work}
Although there are early approaches of visualizing graphs in virtual reality~\cite{Yang:2006}, the technology today has drastically changed.
As VR devices are entering the mainstream, these topics are revisited, e.g., focusing on general virtual reality graph exploration~\cite{vrexp18} or interfaces~\cite{vrinterface18}.
In contrast to those, we designed TGView3D specifically for theory graphs, which have particular structure and whose users have special information needs.
Theory graphs contain multiple types of directed edges and nodes that are rich in information, even seen individually.
Since existing layout algorithms and graph tools do not handle such graphs sufficiently, we contribute custom solutions.

\paragraph{Acknowledgments}
The authors gratefully acknowledge financial support from the OpenDreamKit Horizon 2020 project (\#676541) and the DFG (project OAF; KO 2428/13-1, RA-18723-1).
Furthermore, we acknowledge hardware support and very helpful discussions about layout algorithms from Jonas Müller, Roberto Grosso, and Marc Stamminger.


\bigskip
\myparagraph{The TGView3D System}\label{system}
Theory graphs consist of a set of nodes representing theories and edges representing different kinds of morphisms: 
MMT \textbf{imports} are the most prevalent and induce a directed acyclic subgraph, which is important for understanding the primary structure and thus needs to be prioritized in the layout.
MMT \textbf{views} express translations and similar representation theorems; these may introduce cycles or connect very distant theories.
Each node and edge are labeled with their MMT URI for interoperation with other tools.

We import graphs into TGView3D as JSON files~\cite{TGWiki:on}, which are generated by the MMT system.
Based on the type, we show edges with different colors and indicate their direction with a gradient from light to dark.

\paragraph{Rendering}

Unity fully takes care of rendering.
The programmer has to define objects with transformation matrices, which are responsible for position, scale and rotation.
Usually, an object also contains a mesh describing the geometry and a material for the shading. Colliders can be attached to manage collisions.

The main problem for efficient rendering is the number of nodes and edges. 
To reduce the load, we combine all edges into a single object, hide distant labels and use GPU instancing to draw multiple nodes with a single CPU draw call.

\paragraph{Layout}

\begin{figure}
	\centering
	\begin{subfigure}{.55\textwidth}
		\centering
		\includegraphics[height=130pt]{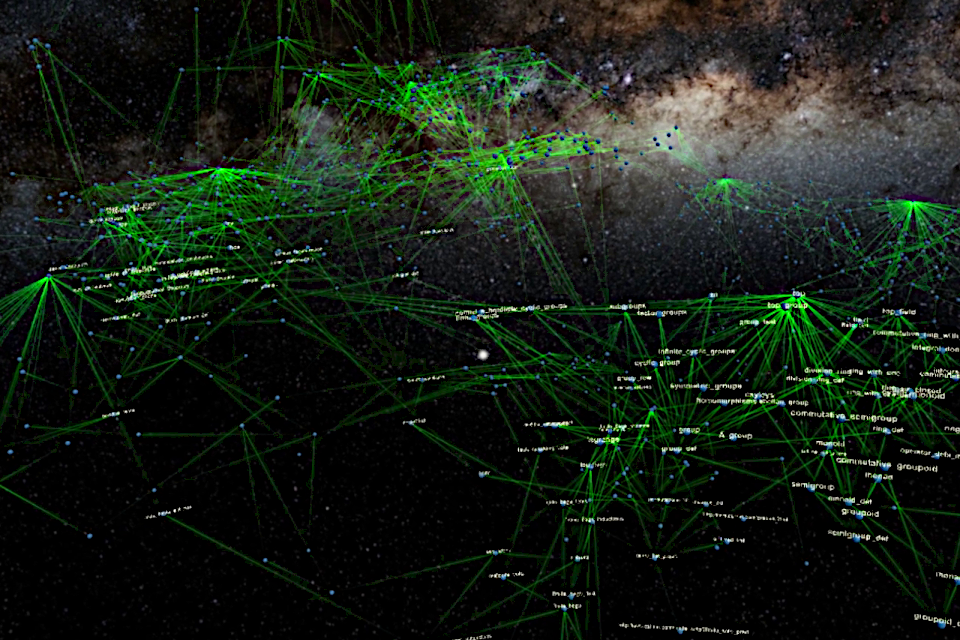}
		\caption{Hierarchical Force-directed Layout}
		\label{fig:nasa_layout}
	\end{subfigure}%
	\hspace{20px}
	\begin{subfigure}{.38\textwidth}
		\centering
		\includegraphics[height=119pt]{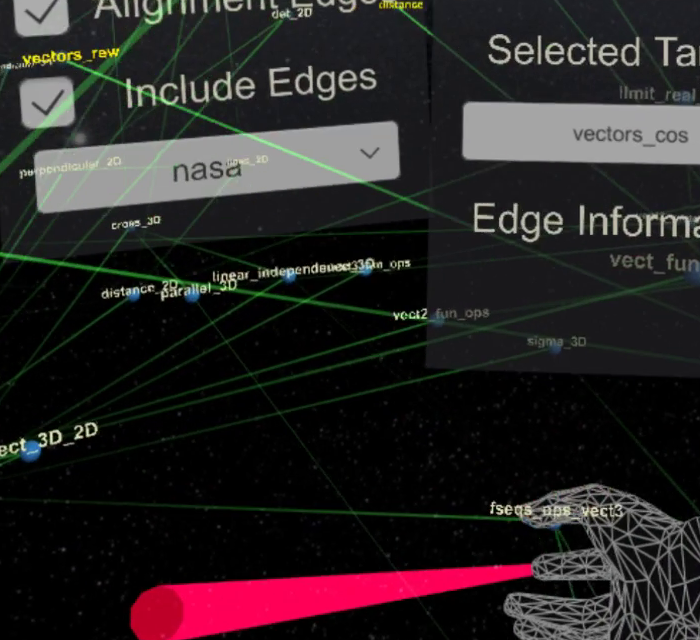}
		\caption{Beam radiating from finger and UI with touchable buttons}
		\label{fig:beamui}
	\end{subfigure}
	\caption{Theory Graph of the PVS Nasa Library with 739 nodes and 2851 edges}
\end{figure}

Our layout algorithm is still experimental; it is based on a force-directed approach: 
ignoring the edge directions, nodes only attract nodes that are connected to them, but repel all of them.
By performing multiple iterations a force balance is attained, generating a layout that forms groups and reduces the average edge-length.
But theory graphs are directed and the induced hierarchy is a central, cognitive aspect for understanding their structure.
Hence, we want to achieve a consistent edge direction going from bottom to top, exploiting the three dimensions.
Therefore, we add a force that repels connected nodes either downwards or upwards, depending on whether they are successors or predecessors.
See Figure~\ref{fig:nasa_layout} for the result of a theory graph layout computation.

\paragraph{User Interaction}

For the purpose of using TGView3D in virtual reality, we have chosen the \textbf{Oculus Rift}.
It is one of the most popular devices and comes with controllers that are designed to translate natural hand movements into the virtual world. 
In order to change settings or start computations, we attach a virtual, touchable UI to the right hand as shown in Figure~\ref{fig:beamui}.
With the feedback from demos, we have developed interactions to explore the graph intuitively.

\paragraph{Graph Exploration}

The first requirement for exploring a graph is the ability to move through the virtual world, which is done with the left control stick.
Rotating and translating the graph are convenient alternatives.
For example, it does not make any semantic difference whether the user pulls the graph closer or flies towards it. 
Scaling has a different objective.
It allows the user to switch between a global and a local view by changing the node spacing.
We achieve this effect by scaling the positions of the nodes instead of the whole graph.

To perform the interactions mentioned above, we use hand gestures in combination with an initial button press.
Rotation is controlled by moving the hand horizontally.
We intentionally limit this to rotations around the vertical axis to keep the hierarchy consistent, i. e., edges pointing downwards will keep this property.
When translating, the graph simply follows the movement of the hand. 
Bringing the controllers closer together scales down the graph, while the opposite happens, when the distance between them is increased.

\paragraph{Node Selection}

\begin{figure}[htb]\centering
  \begin{subfigure}{.45\textwidth}\centering
    \includegraphics[width=1\linewidth]{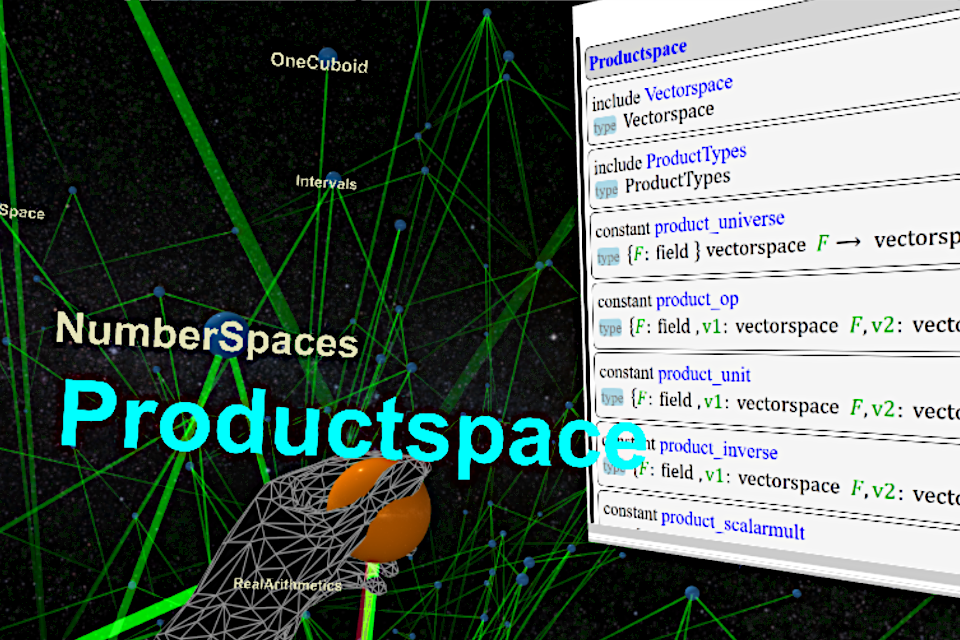}
    \caption{Grabbing and selecting a node}\label{fig:tractor}
  \end{subfigure}%
  \hspace{20px}
  \begin{subfigure}{.45\textwidth}\centering
    \includegraphics[width=1\linewidth]{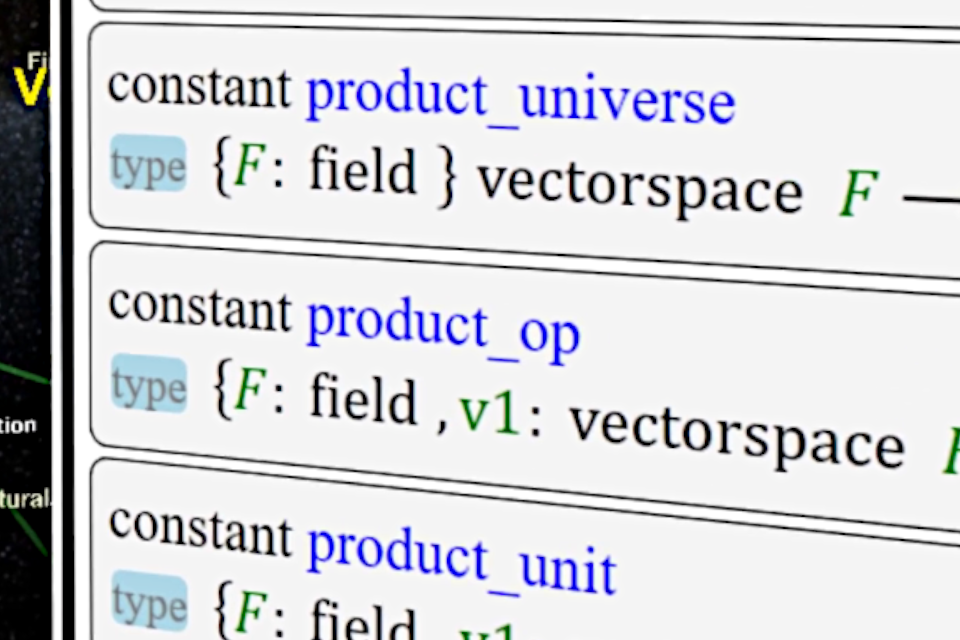}
    \caption{Virtual screen with node details}\label{fig:closeup}
  \end{subfigure}
  \caption{Interacting with nodes of the MitM Ontology theory graph}
  \label{fig:mitm}
\end{figure}

Once the user has reached an interesting part of the graph, he might want to gain further insights regarding certain nodes.
To select them, the user can pick up nodes by performing a grab motion or pull distant ones closer with the help of a tractor beam, depicted in Figure~\ref{fig:beamui}.
The labels of the selected node and one connected node appear in the UI and the user can iterate through the latter with the right control stick.
Touching the labels opens a website in the browser that contains information about the respective node, which we then stream into the application as a virtual touch screen, shown in Figure~\ref{fig:mitm}.

\paragraph{Information Filtering}

Although we now have introduced exploration tools, complex graphs often contain so much information that it is hard to recognize structure.
Exploiting the third dimension already helps by creating more space for the graph, but the parts in the distance still interfere with the part in focus.
To prevent this, the user can hide distant parts of the graph.
We achieve more sophisticated effects by only showing a subset of edges: the user can 

\begin{itemize}
\item disable edges of certain types and, optionally, reload the graph.
  This does not only reduce the number of edges, but also grants new insights how certain combinations of edge types influence the layout.
\item hide all edges that do not belong to the reachable subgraph of a selected node or the opposite subgraph, using reversed edge directions.
  The volume occupied by the remaining edges indicates the relation to the original graph.
\item focus on the neighborhood of selected nodes by hiding all others.
\end{itemize}



\bigskip
\myparagraph{Conclusion and Future Work}\label{sec:conc}
We have presented an interactive 3D theory graph viewer, that can be integrated into the everyday workflow of scientists.
First, our layout algorithm makes use of the third spatial dimension to project the hierarchy of theory graphs, which is difficult to emulate in 2D.
By implementing methods to explore the resulting structures globally and locally, we have broken up the information density of theory graphs.
Finally, the VR implementation avoids non-intuitive mouse and keyboard 3D navigation, but still allows to access insights from the web.

While TGView3D can be run on an integrated graphics processor, the Oculus Rift, has strict system requirements~\cite{ORRequirements:on}.
To reach a wide audience of users, we have already built versions that do not require any VR devices for multiple platforms, including a provisional web variant~\cite{TGW3DWEB:on}.

Future work will focus on extending the functionalities of TGView3D, especially with regard to integration of use case--specific interactions and tools such as
proof assistants or library management tools like refactoring or search.

\printbibliography
\end{document}